\documentclass[lettersize,journal]{IEEEtran}
\IEEEoverridecommandlockouts
\usepackage[utf8]{inputenc}
\usepackage{cite}
\usepackage{amsmath,amssymb,amsfonts}
\usepackage{algorithm} 
\usepackage{graphicx}
\usepackage{caption}
\usepackage{subcaption}
\usepackage{textcomp}
\usepackage{xcolor}
\usepackage{array}
\usepackage{textcomp}
\usepackage{stfloats}
\usepackage{url}
\usepackage{verbatim}
\usepackage{amsthm, bm, hyperref, mathrsfs, indentfirst, longtable}
\usepackage{svg}
\usepackage{longtable}
\usepackage{tabularx}
\usepackage{makecell}
\usepackage{booktabs,float}
\usepackage{bbm}
\usepackage{balance}
\usepackage{threeparttable}
\usepackage{multirow}
\usepackage{booktabs}
\usepackage{caption}
\usepackage{float}
\usepackage{booktabs}
\usepackage{stfloats}

\def\BibTeX{{\rm B\kern-.05em{\sc i\kern-.025em b}\kern-.08em
    T\kern-.1667em\lower.7ex\hbox{E}\kern-.125emX}}
\begin{document}

\title{Multimodal Radio and Vision Fusion for Robust Localization in Urban V2I Communications}

\author{\IEEEauthorblockN{Can Zheng\IEEEauthorrefmark{1}, Jiguang He\IEEEauthorrefmark{2}, Chung G. Kang\IEEEauthorrefmark{1}, Guofa Cai\IEEEauthorrefmark{3}, Henk Wymeersch\IEEEauthorrefmark{4}}\\
\IEEEauthorblockA{\IEEEauthorrefmark{1}School of Electrical Engineering, Korea University, Seoul, Republic of Korea}\\
\IEEEauthorblockA{\IEEEauthorrefmark{2}School of Computing and Information Technology, Great Bay University, Dongguan 523000, China}\\
\IEEEauthorblockA{\IEEEauthorrefmark{3}School of Information Engineering, Guangdong University of Technology, Guangzhou, China}\\
\IEEEauthorblockA{\IEEEauthorrefmark{4}Department of Electrical Engineering, Chalmers University of Technology, Gothenburg, Sweden}
}

\maketitle
\begin{abstract}
    Accurate localization is critical for vehicle-to-infrastructure (V2I) communication systems, especially in urban areas where GPS signals are often obstructed by tall buildings, leading to significant positioning errors, necessitating alternative or complementary techniques for reliable and precise positioning in applications like autonomous driving and smart city infrastructure. This paper proposes a multimodal contrastive learning regression based localization framework for V2I scenarios that combines channel state information (CSI)  with visual information to achieve improved accuracy and reliability. The approach leverages the complementary strengths of wireless and visual data to overcome the limitations of traditional localization methods, offering a robust solution for V2I applications. Simulation results demonstrate that the proposed CSI and vision fusion model significantly outperforms traditional methods and single modal models, achieving superior localization accuracy and precision in complex urban environments.
\end{abstract}


\section{Introduction}

    Accurate localization is a cornerstone of vehicle-to-infrastructure (V$2$I) communication systems, particularly in urban environments where the global positioning system (GPS) often fails due to signal obstructions from tall buildings, resulting in significant positioning errors \cite{V2I_loc}. In such scenarios, traditional GPS-based localization methods are inadequate, requiring alternative or complementary techniques for reliable and precise positioning in applications like autonomous driving and smart city infrastructure.

    The evolution of wireless localization from $4$G to $5$G networks highlights the increasing demand for high-precision positioning. In $4$G networks, localization techniques such as observed time difference of arrival (OTDOA) and uplink time difference of arrival (UL-TDOA) offered accuracy in the tens of meters, mainly for emergency services \cite{4g_loc}. The advent of $5$G networks has marked a significant leap forward, introducing dedicated positioning reference signals (RSs) and advanced techniques such as round trip time (RTT) and angle-based methods \cite{5G_loc}. These advancements enable positioning accuracies down to centimeters, driven by wider bandwidths, massive antenna arrays, and enhanced signal processing capabilities.

    \begin{figure}[t]
        \centering
        \captionsetup{font=footnotesize}
        \begin{subfigure}[t]{0.48\columnwidth} 
            \centering
            \includegraphics[width=\textwidth]{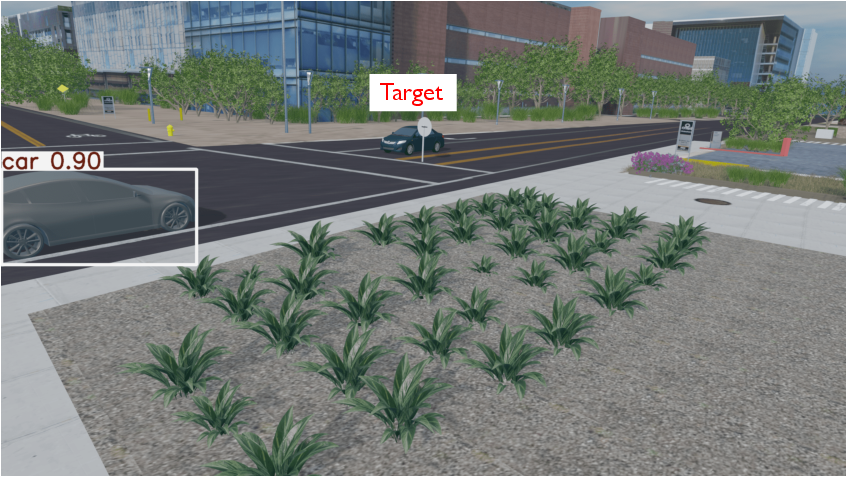} 
            \caption{Miss detection of the target.}
        \end{subfigure}
        \hfill 
        \begin{subfigure}[t]{0.48\columnwidth}
            \centering
            \includegraphics[width=\textwidth]{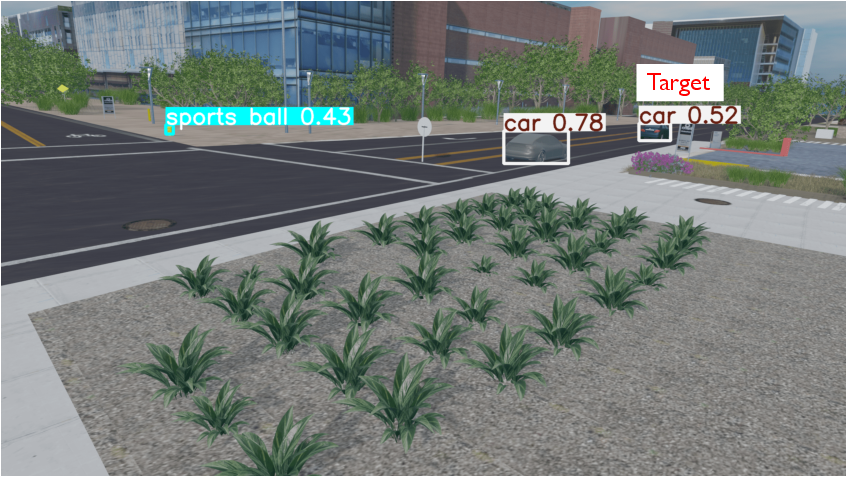} 
            \caption{False positive of the target.}
        \end{subfigure}
        \caption{Two common failure modes observed in a YOLO-based object detection system, highlighting the inherent limitations of relying solely on visual input for accurate localization.}
        \label{fig:vision_miss}
    \end{figure}

    Despite the potential of V$2$I communication for localization, several challenges persist. V$2$I systems rely on interactions between vehicles and roadside units (RSUs), but multipath propagation, non-line-of-sight (NLOS) conditions, and the nonlinear, non-Gaussian nature of wireless measurements can severely degrade performance \cite{V2I_2}. Traditional angle-of-arrival (AoA) estimation methods, such as the multiple signal classification (MUSIC) algorithm, struggle with multipath interference due to the rank deficiency of the covariance matrix \cite{MUSIC}. Techniques like spatial smoothing can mitigate some of these issues, but their effectiveness diminishes when the line-of-sight (LoS) path is weak \cite{Spatial_Smoothing}. Furthermore, range estimation accuracy is closely tied to system bandwidth. In narrowband systems, achieving high precision requires substantial RS overhead, which is resource-intensive and inefficient. These issues highlight the need for innovative approaches to improve localization accuracy in complex urban environments. 
    
    To address the limitations of current localization systems, integrating  channel state information (CSI) with visual data from RSU-mounted cameras offers a promising solution. Visual data provides positional cues from environmental landmarks, unaffected by wireless signal distortions, while CSI ensures stability when visual data is unreliable. (see Fig. \ref{fig:vision_miss}). 
    Fusing these modalities creates a robust, accurate localization system for complex environments. Current research has explored various avenues to enhance localization in challenging environments. For instance, one approach leverages beam received power and visual data for GPS denoising \cite{GPSdenoise}, aiming to refine positioning in scenarios where GPS signals are unreliable. Another notable effort involves a meta-learning framework for joint localization using multiple cameras and RSUs \cite{CV+CSI}, showcasing the potential of advanced machine learning techniques in complex multi-sensor systems. 
    
    In this work, we propose a multimodal contrastive learning regression based localization framework for V$2$I scenarios that combines CSI with visual information to achieve improved accuracy and reliability. Our main contributions are: (i) proposing a novel localization framework that utilizes only a single RSU with a monocular camera to achieve high-precision positioning, eliminating the need for multiple RSUs; (ii) developing an innovative approach for feature alignment and fusion between CSI and visual modalities, enabling complementary multimodal integration for robust localization.
    This distinct approach aims to demonstrate robust localization performance even in highly constrained sensing environments.
    \begin{figure}[t]
        \centering
        \captionsetup{font=footnotesize}
        \includegraphics[width=0.8\linewidth]{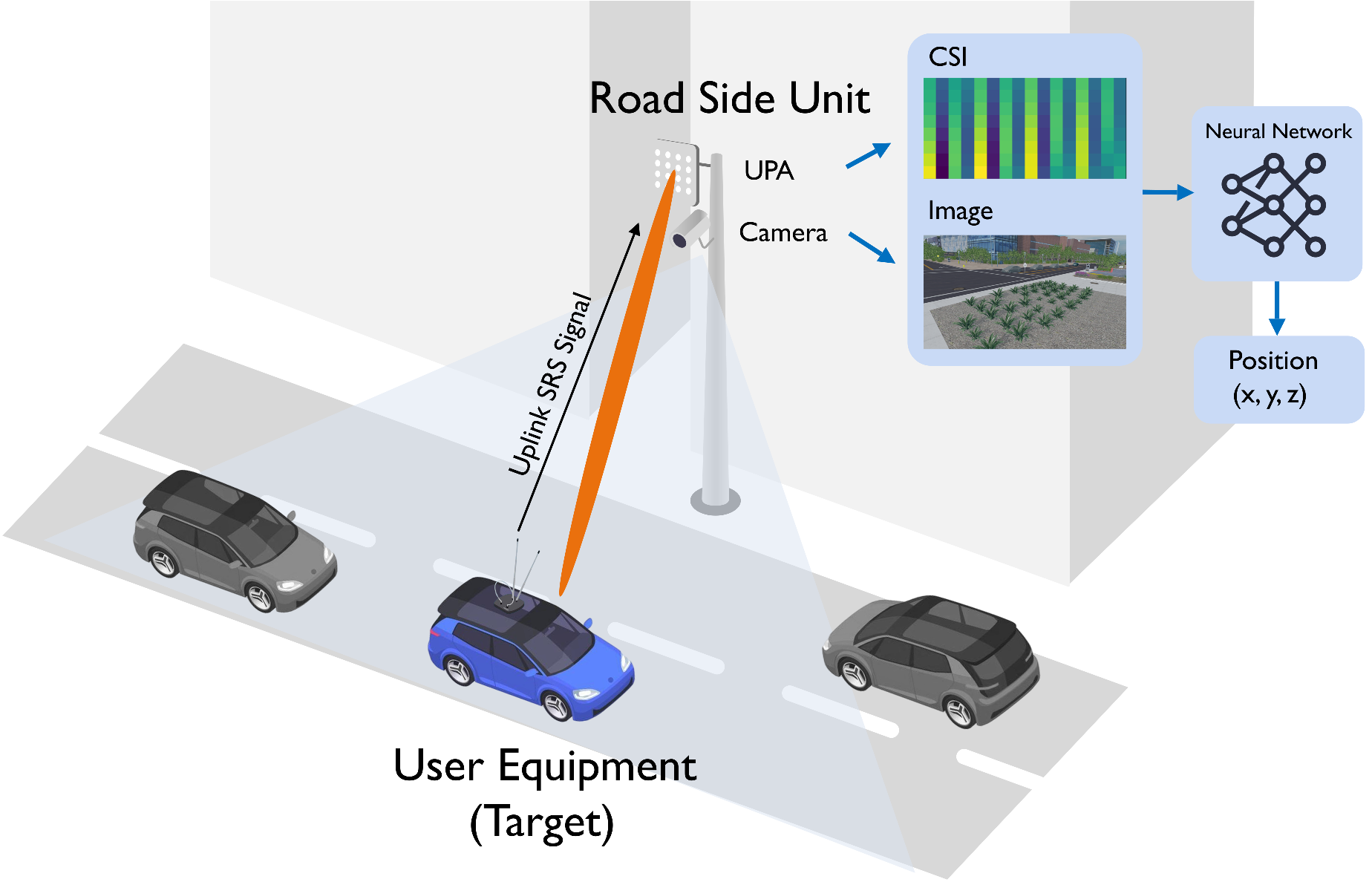}
        \caption{Demonstration of considered urban V$2$I scenario. The collected CSI and the image data in RSU are fed into a neural network to estimate the $3$D position of the UE.}
        \label{fig:scenario}
    \end{figure}

    \textit{Notations}: Bold lowercase letters denote vectors (e.g., $\mathbf{x}$), and bold uppercase letters denote matrices or tensors (e.g., $\mathbf{X}$). $\text{diag}(\mathbf{x})$ denotes a square diagonal matrix with entries of $\mathbf{X}$ on its diagonal. $[\mathbf{X}]_{a,:}$ and $[\mathbf{X}]_{:,b}$ denote the $a$-th row, $b$-th column vectors of $\mathbf{X}$. The superscripts $(\cdot)^\mathsf{T}$ and $(\cdot)^\mathsf{H}$ represent the transpose and Hermitian (conjugate transpose) operations, respectively. The operator $[\cdot;\cdot]$ denotes the concatenation operation, while $\|\cdot\|_2$ denotes the Euclidean norm of a vector and $|\cdot|$ denotes element-wise magnitude of a vector or a matrix. $\otimes$ denotes the Kronecker product. Unless otherwise specified, $\mathbb{R}$ and $\mathbb{C}$ denote the sets of real and complex numbers, respectively. $\mathrm{Cov}(\cdot,\cdot)$ denotes the covariance.

\section{System Model}
    In this section, we delve into the system model considered, as shown in Fig. \ref{fig:scenario}. The RSU integrated with monocular cameras and other sensors use uplink (UL) CSI from the UE and sensing data such as visual data to enable high-precision localization. Meanwhile, the RSU employs orthogonal frequency division multiplexing (OFDM) with $K$ subcarriers and is equipped with a uniform planar array (UPA) comprising $M$ antenna elements. We aim to estimate the UE position $\mathbf{p}_\text{UE} = [p_\text{UE}^x, p_\text{UE}^y, p_\text{UE}^z]^\mathsf{T}$ in a multipath environment.

    In practical deployments, the RSU can instruct the UE to transmit an RS, enabling UL channel estimation. Let $\hat{h}_{m,k}$ represent the channel response associated with the $m$-th antenna and $k$-th subcarrier. It can be modeled as the product of an amplitude response $|\hat{h}_{m,k}|$ and a phase response $\angle\hat{h}_{m,k}$, expressed as a summation over multiple propagation paths. Specifically, the channel gain can be written as
    \begin{align}
            \hat{h}_{m,k}  =|\hat{h}_{m,k}|e^{\angle\hat{h}_{m,k}}=\sum_{l=1}^{L} \alpha_l e^{-j 2\pi (f_c + k\Delta f) \tau_l} {a}_m(\theta_l, \phi_l),
    \end{align} 
    where $f_c$ denotes the carrier frequency and $L$ is the total number of propagation paths; for the $l$-th path, $\alpha_l$, $\tau_l$, and ${a}_m(\theta_l, \phi_l)$ denote the complex gain, the time of arrival (ToA), and the $m$-th element of the corresponding steering vector, respectively. Here, $\theta_l$ represents the elevation angle and $\phi_l$ represents the azimuth angle for the $l$-th path's direction of arrival. The term $\Delta f$ represents the subcarrier spacing.
    Let $\hat{\mathbf{H}}_\text{F} \in \mathbb{C}^{M\times K}$ represent the sampled frequency-domain CSI matrix. We can then separate the matrix into an outer product of array steering vector $\mathbf{a}(\phi_l, \theta_l)\in \mathbb{C}^{M\times1}$ and delay steering vector $\mathbf{d}(\tau_l)\in \mathbb{C}^{K\times1}$ for path $l$:
    \begin{align}
        \mathbf{a}(\theta_l, \phi_l) = \begin{bmatrix}
        a_1(\theta_l, \phi_l) \\
        a_2(\theta_l, \phi_l) \\
        \vdots \\
        a_M(\theta_l, \phi_l)
        \end{bmatrix}, \ \mathbf{d}(\tau_l) = \begin{bmatrix}
        e^{-j 2\pi f_c \tau_l} \\
        e^{-j 2\pi (f_c + 1\Delta f) \tau_l} \\
        \vdots \\
        e^{-j 2\pi (f_c + (K-1)\Delta f) \tau_l}
        \end{bmatrix}.
    \end{align}


    Assuming no mutual coupling between antenna elements, we consider a UPA with $M_x$ elements along the $x$-axis and $M_y$ elements along the $y$-axis, such that $M = M_x M_y$. 
    Let $\mathbf{a}_{x}(\theta_l, \phi_l)$ and $\mathbf{a}_{y}(\theta_l, \phi_l)$ denote the steering vectors along the $x$-axis and $y$-axis, respectively, given by
        \begin{align}
            \mathbf{a}_{x}(\theta_l, \phi_l) &= \left[1, e^{j 2\pi\frac{d}{\lambda}\cos\phi_l \sin\theta_l},\cdots, e^{j 2\pi\frac{(M_x-1)d}{\lambda}\cos\phi_l \sin\theta_l}\right]^\mathsf{T},\\
            \mathbf{a}_{y}(\theta_l, \phi_l) &= \left[1, e^{j 2\pi\frac{d}{\lambda}\sin\phi_l \sin\theta_l},\cdots, e^{j 2\pi\frac{(M_y-1)d}{\lambda}\sin\phi_l \sin\theta_l}\right]^\mathsf{T},
        \end{align}
    where $d$ denotes the spacing between adjacent antenna elements, typically set to half the wavelength. Finally, the steering vector for the $l$-th path, denoted as $\mathbf{a}(\theta_l, \phi_l)$, is given by $\mathbf{a}(\theta_l, \phi_l) = \mathbf{a}_{y}(\theta_l, \phi_l) \otimes \mathbf{a}_{x}(\theta_l, \phi_l)$, and the CSI matrix is $\hat{\mathbf{H}}_\text{F} = \sum_{l=1}^{L} \alpha_l \mathbf{a}(\theta_l, \phi_l) \mathbf{d}^\mathsf{T}(\tau_l)$.



\section{$3$D Localization Methods}
    This section briefly introduces the traditional localization method based on RS, followed by the proposed multimodal CSI-vision fusion framework.
    \subsection{Traditional Method}
    \label{sec:traditional}
    \subsubsection{Angle Estimation}
     For each subcarrier, we have a single measurement. The traditional method, based on the strongest LoS, treats each single measurement within a subcarrier as an independent snapshot. This means that for each subcarrier $k$, we effectively have one realization of the channel $[\hat{\mathbf{H}}_\text{F}]_{:,k}$. Since the true covariance matrix is not available in practical scenarios, we approximate it using the sample mean covariance matrix:
    \begin{align}
        \hat{\mathbf{R}}_h  =\frac{1}{K} \sum_{k=1}^K [\hat{\mathbf{H}}_\text{F}]_{:,k}[\hat{\mathbf{H}}_\text{F}]_{:,k}^\mathsf{H}=\frac{1}{K} \hat{\mathbf{H}}_\text{F} \hat{\mathbf{H}}_\text{F}^\mathsf{H}.
    \end{align}
    
    The eigenvalue decomposition (EVD) of $\hat{\mathbf{R}}_h$ is given by
    \begin{align}
        \hat{\mathbf{R}}_h = \mathbf{U} \boldsymbol{\Lambda} \mathbf{U}^\mathsf{H} = \mathbf{U}_s \boldsymbol{\Lambda}_s \mathbf{U}_s^\mathsf{H} + \mathbf{U}_n \boldsymbol{\Lambda}_n \mathbf{U}_n^\mathsf{H},
    \end{align}
    where $\boldsymbol{\Lambda} = \mathrm{diag}(\lambda_1, \dots, \lambda_M)$ is the diagonal matrix of eigenvalues $\lambda_i$ of $\hat{\mathbf{R}}_h$, sorted such that $\lambda_1 \geq \lambda_2 \geq \cdots \geq \lambda_M \geq 0$. The matrices $\mathbf{U}_s$ and $\mathbf{U}_n$ consist of the eigenvectors corresponding to the signal and noise subspaces, respectively, with $\boldsymbol{\Lambda}_s$ and $\boldsymbol{\Lambda}_n$ denoting the associated diagonal eigenvalue matrices.
     In our localization model, we only account for LoS propagation. Under this assumption, $\mathbf{U}_s = [\mathbf{U}_s]_{:,1}$ and $\mathbf{\Lambda}_s = [\lambda_1]$. The noise subspace $\mathbf{U}_n$ is then composed of the remaining $M-1$ eigenvectors associated with the smaller eigenvalues $\lambda_2, \dots, \lambda_M$. Subsequently, the MUSIC pseudo-spectrum is computed as follows \cite{MUSIC}:
        \begin{align}
            P_{\text{MUSIC}}(\theta, \phi) = \frac{1}{\mathbf{a}^\mathsf{H}(\theta, \phi) \mathbf{U}_n \mathbf{U}_n^\mathsf{H} \mathbf{a}(\theta, \phi) }.
        \end{align}
        
    The angles estimates are then obtained by locating the peak of the pseudo-spectrum as follows:
        \begin{align}
            (\hat{\theta}, \hat{\phi}) = \arg\max_{(\theta, \phi)} P_{\text{MUSIC}}(\theta, \phi),
        \end{align}
    evaluated over a predefined grid of $(\theta, \phi)$ values.

    \subsubsection{Ranging}
    To estimate the ToA for range estimation, we first apply an inverse fast Fourier transform (IFFT) to $\hat{\mathbf{H}}_\text{F}$ along the frequency domain to obtain the delay-domain response:
    \begin{align}
    \hat{\mathbf{H}}_\text{D} = \text{IFFT}(\hat{\mathbf{H}}_\text{F}) \in \mathbb{C}^{M \times K},
    \end{align}
    where each row of $\hat{\mathbf{H}}_\text{D}$ represents the delay-domain channel impulse response at a single antenna. We compute the spatial average over all $M$ antennas:
    \begin{align}
    \bar{\mathbf{h}}(\tau) = \frac{1}{M} \sum_{m=1}^M \left[\hat{\mathbf{H}}_{\text{D}}\right]_{m,:}(\tau).
    \end{align}
    
    The ToA is then estimated by locating the peak of the averaged impulse response:
    \begin{align}
    \hat{\tau} = \arg \max_t |\bar{\mathbf{h}}(\tau)|.
    \end{align}
    
    Assuming perfect synchronization between the UE and RSU, the estimated distance 
    $\hat{d} = \hat{\tau} c$ is calculated, where $c$ is the speed of light.
    \subsubsection{Positioning}
    With the estimated angles $(\hat{\theta},\hat{\phi})$ and calculated distance $\hat{d}$, the estimated position of the UE $\hat{\mathbf{p}}_\text{UE}$ can be determined as:
    \begin{align}
        \hat{\mathbf{p}}_\text{UE} = \hat{d} \cdot \hat{\mathbf{r}}(\hat{\theta},\hat{\phi}) + \mathbf{p}_\text{RSU},
    \end{align}
    where $\hat{\mathbf{r}}(\hat{\theta},\hat{\phi}) = [\sin \hat{\theta}\cos\hat{\phi}, \sin \hat{\theta}\sin \hat{\phi}, \cos{\hat{\theta}}]^\mathsf{T}$ is the unit vector of arrival direction, and $\mathbf{p}_\text{RSU} = [p_\text{RSU}^x, p_\text{RSU}^y, p_\text{RSU}^z]^\mathsf{T}$ is the known RSU location.

    \subsection{Proposed Multimodal Localization via CSI and Visual Feature Fusion}
    \label{sec:CSI+Vision}
    \begin{figure}[t]
        \centering
        \captionsetup{font=footnotesize}
        \includegraphics[width=\linewidth]{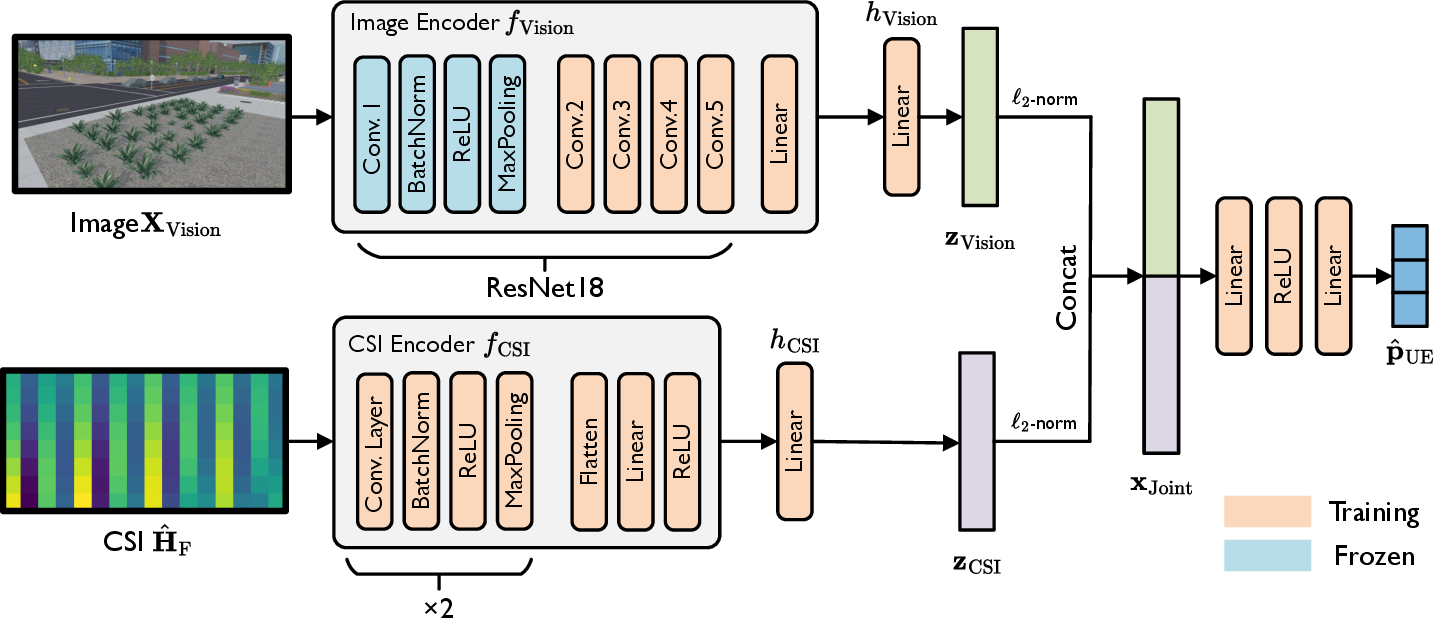}
        \caption{The proposed multimodal localization model framework.}
        \label{fig:framework}
    \end{figure}
    In this section, we introduce the proposed localization framework, as shown in Fig. \ref{fig:framework}. Specifically, given that the CSI and the corresponding visual data are collected at the same physical location, this motivates us to learn cross-modal aligned representations for localization. 
    
    \subsubsection{Feature Extraction and Encoding}
    In this step, both the raw CSI and visual data are processed to extract their respective features. The CSI data is primarily used for its amplitude response, while the visual data is used to extract the corresponding image features. 
    
    \textbf{CSI encoder} $f_\text{CSI}(\cdot)$: The raw CSI is processed to extract its amplitude response, which is then encoded into a low-dimensional embedding vector. 
    Given the amplitude of the CSI matrix $|\hat{\mathbf{H}}_\text{F}| \in \mathbb{R}^{M \times K}$, the CSI feature vector $\mathbf{z}_\text{CSI} \in \mathbb{R}^{d_\text{CSI}}$ is extracted as
\begin{align}
    \mathbf{z}_\text{CSI} = f_\text{CSI}\left(|\hat{\mathbf{H}}_\text{F}|\right),
\end{align}
where $d_\text{CSI}$ denotes the dimension of the CSI feature.

    \textbf{Vision encoder} $f_\text{Vision}(\cdot)$: The input image data is processed using a fine-tuned ResNet-$18$ to extract visual features \cite{resnet}. Given an image $\mathbf{X}_\text{Vision} \in \mathbb{R}^{W_\text{Vision} \times H_\text{Vision} \times C_\text{Vision}}$, where $W_\text{Vision}$, $H_\text{Vision}$, and $C_\text{Vision}$ represent the spatial width, height, and RGB/IR channels, respectively, the visual feature vector $\mathbf{z}_\text{Vision} \in \mathbb{R}^{d_\text{Vision}}$ is obtained as
    \begin{align}
        \mathbf{z}_\text{Vision} = f_\text{Vision}\left(\mathbf{X}_\text{Vision}\right),
    \end{align}
    where $d_\text{Vision}$ denotes the dimension of the vision feature.

    \subsubsection{CSI-Vision Alignment}

    After feature extraction, we project CSI and vision inputs into a shared embedding space, aligning features from the same location across modalities to enable effective cross-modal localization and representation learning \cite{CLIP}. Two projection heads are introduced to achieve this alignment:

    \textbf{CSI projection head:} The projection head $h_\text{CSI}(\cdot)$ maps the CSI feature $\mathbf{z}_\text{CSI}$ into the shared embedding space:
\begin{align}
    \mathbf{x}_\text{CSI} = h_\text{CSI}(\mathbf{z}_\text{CSI}),
\end{align}
where $\mathbf{x}_\text{CSI} \in \mathbb{R}^{d_\text{Embed}}$ denotes the projected CSI embedding, and $d_\text{Embed}$ is the dimensionality of the shared embedding space.
    
    \textbf{Vision projection head:} The projection head $h_\text{Vision}(\cdot)$ maps the vision feature $\mathbf{z}_\text{Vision}$ into the shared embedding space:
\begin{align}
    \mathbf{x}_\text{Vision} = h_\text{Vision}(\mathbf{z}_\text{Vision}),
\end{align}
where $\mathbf{x}_\text{Vision} \in \mathbb{R}^{d_\text{Embed}}$ denotes the projected vision embedding in the shared space.

To enhance numerical stability and facilitate similarity computation, we apply $\ell_2$-normalization to the outputs of both projection heads, ensuring that the resulting embeddings have unit norm. This results in the normalized feature vectors $\bar{\mathbf{x}}_\text{CSI}$ and $\bar{\mathbf{x}}_\text{Vision}$:
\begin{align}
    \bar{\mathbf{x}}_\text{CSI} = \frac{\mathbf{x}_\text{CSI}}{\Vert \mathbf{x}_\text{CSI} \Vert_2}, \quad
    \bar{\mathbf{x}}_\text{Vision} = \frac{\mathbf{x}_\text{Vision}}{\Vert \mathbf{x}_\text{Vision} \Vert_2}.
\end{align}
    
    \subsubsection{Localization Regression}
    After aligning CSI and visual features in the shared embedding space, we fuse them through concatenation along the feature dimension to form a joint feature vector: 
    \begin{align}
    \mathbf{x}_\text{Joint} = [\bar{\mathbf{x}}_\text{CSI}; \bar{\mathbf{x}}_\text{Vision}].
    \end{align}
    
    This vector is fed into a regression network to predict the UE’s $3$D position:
    \begin{align}
        \hat{\mathbf{p}}_\text{UE} = f_\text{Predict}(\mathbf{x}_\text{Joint}),
    \end{align}
    where $f_\text{Predict}(\cdot)$ represents the prediction head.
    
    \subsubsection{Learning Objective of the Neural Network}
    Our framework is designed to jointly optimize cross-modal feature alignment and accurate UE localization through a unified, adaptively weighted total loss function. 

    \textbf{Localization loss:} The objective of the localization loss is to directly optimize the accuracy of the predicted $3$D coordinates with respect to the ground-truth location. We adopt the mean squared error (MSE) as the localization loss function. The localization loss among the batch is defined as
    \begin{align}
        \mathcal{L}_{L} = \frac{1}{B}\sum_{i=1}^B \Vert \mathbf{p}_\text{UE}^{(i)} - \hat{\mathbf{p}}_\text{UE}^{(i)}\Vert^2_2,
    \end{align}
    where $\mathbf{p}_\text{UE}^{(i)}$ and $\hat{\mathbf{p}}_\text{UE}^{(i)}$ represent the ground truth location and the estimated location for the $i$-th sample, respectively, and $B$ denotes the batch size.

    \textbf{Contrastive loss:} The objective of the contrastive loss is to encourage the CSI and visual features extracted from the same physical location to be close in the shared embedding space, while ensuring that features originating from different locations are well separated. For a batch of $B$ samples, let \(\bar{\mathbf{x}}_{\text{CSI},i} \in \mathbb{R}^{d_\text{Embed}}\) and \(\bar{\mathbf{x}}_{\text{Vision},j} \in \mathbb{R}^{d_\text{Embed}}\) denote the \(\ell_2\)-normalized embeddings of the $i$-th CSI features and the $j$-th vision features, respectively. We construct a similarity matrix \(\mathbf{S} \in \mathbb{R}^{B \times B}\), where the $(i,j)$-th entry is given by $S_{ij} = \frac{\bar{\mathbf{x}}_{\text{CSI},i}^\mathsf{T} \bar{\mathbf{x}}_{\text{Vision},j}}{\alpha}$, and $\alpha > 0$ is a temperature scaling parameter that controls the sharpness of the similarity distribution.
    
    Next, we define the contrastive loss using a cross-entropy loss function. In this batch, we expect each CSI embedding $\bar{\mathbf{x}}_{\text{CSI},i}$ to have the highest similarity with its corresponding wireless signal embedding $\bar{\mathbf{x}}_{\text{Vision},i}$. Therefore, the indices on the diagonal of the matrix are used as the positive samples in the labels. The contrastive loss is defined as

    \begin{align}
    \mathcal{L}_C = -\frac{1}{B} \sum_{i=1}^{B} \log \left( \frac{\exp(S_{ii})}{\sum_{j=1}^{B} \exp(S_{ij})} \right).
    \end{align}

    \textbf{Adaptive weighted total loss:}
    To effectively balance the importance of localization regression and contrastive learning, we adopt an adaptive weighting strategy based on task uncertainty \cite{loss}. Let $s_{L} = \log (\sigma_L^2)$ and $s_{C}=\log (\sigma_C^2)$ denote the learnable log-variances for $\mathcal{L}_{L}$ and $\mathcal{L}_C$, respectively. The total loss is formulated as
    \begin{align}
        \mathcal{L}_{\text{Total}} = \frac{1}{2\exp(s_C)} \mathcal{L}_C + \frac{1}{2\exp(s_L)} \mathcal{L}_L + \frac{1}{2}(s_C + s_L).
    \end{align}
    
    Here, the first and second terms act as inverse-variance weights, dynamically balancing each task's contribution. The third term prevents overly low uncertainty to avoid overconfidence. By minimizing $\mathcal{L}_{\text{Total}}$, the network not only learns to jointly optimize localization regression and contrastive alignment but also automatically adjusts the task importance based on the underlying uncertainty.
    
\section{Simulation Results}
    \subsection{Simulation Settings}

    We utilize the DeepVerse $6$G dataset, specifically the DT$31$ scenario, a digital twin of DeepSense $6$G Scenario $31$ \cite{DeepVerse, DeepSense}. This scenario comprises a total of $7012$ samples, which we divide into a $70\%$ training set and a $30\%$ test set. This scenario captures an outdoor urban wireless environment along a two-way city street with moderate vehicular traffic, offering realistic channel and sensing conditions. In the dataset, the target UE for to be localized is marked in green, while other vehicles are grayed out. We choose DeepVerse $6$G over DeepSense $6$G because the latter does not provide CSI, which is a critical requirement for our localization framework. DeepVerse $6$G generates CSI using ray tracing, enabling realistic simulation of wireless propagation \cite{Remcom}. Accordingly, in our simulations, we assume that perfect CSI is available at the RSU.
    
    The detailed parameter settings are listed in Table \ref{tab:param}.
    \subsection{Evaluated Methods}
    We evaluate the performance of the proposed method:
    \begin{enumerate}
        \item \textbf{CSI$+$Vision model:} The method discussed in Sec. \ref{sec:CSI+Vision}.
        \item \textbf{CSI only model:} Use residual blocks and use CSI data only for localization. 
        \item \textbf{Vision only model:} Use ResNet-$18$ and use vision data only for localization. Note that this method requires \textit{a priori} information about the target vehicle (e.g., color or approximate location.
        \item \textbf{Traditional method:} The method discussed in Sec. \ref{sec:traditional}.
    \end{enumerate}

    The detailed parameter settings are listed in Table \ref{tab:modelconfig}.
    
    \begin{table}[t]
    \centering
    \caption{Parameter settings.}
    \label{tab:param}
    \begin{tabular}{ll} 
        \toprule
        \textbf{Parameters} & \textbf{Value} \\
        \midrule 
        Number of antennas ($M$ ($M_x\times M_y$)) & $16\ (4\times 4)$\\
        Number of subcarriers ($K$) & $8$\\
        BS carrier frequency and bandwidth & $60$ GHz, $50$ MHz\\
        Number of paths ($L$) & $25$\\
        Speed of light ($c$) & $3\times 10^8 $m/s\\
        Image size ($W_I, H_I, C_I$) & $960, 540, 3$\\
        Feature sizes ($d_\text{CSI}$, $d_\text{Vision}$) & $64$\\
        Embedding size ($d_\text{Embed}$) & $16$\\
        Batch size ($B$) & $32$\\
        Temperature parameter ($\alpha$) & $0.07$\\
        Training epochs & $50$\\
        Initial learning rate & $0.001$\\
        Decay mode& $0.5$ per $10$ epochs\\
        \bottomrule
    \end{tabular}
    \end{table}

    \begin{table}[t]
    \centering
    \caption{Architectures for considering models.}
    \begin{tabular}{|l|l|l|l|}
    \hline
    \textbf{Module} & \textbf{CSI+Vision} & \textbf{CSI Only} & \textbf{Vision Only} \\
    \hline
    $f(\cdot)$ & \makecell[l]{\textit{CSI:}\\ Conv$2$d($1$,$32$),\\ Conv$2$d($32$,$64$),\\ Linear($256$,$64$),\\ Linear($64$,$16$);\\\textit{Vision:}\\ResNet-$18$,\\ FC($512$,$64$)} & \makecell[l]{Conv$2$d($2$,$16$),\\ ResBlock($16$,$32$),\\ ResBlock($32$,$64$),\\ ResBlock($64$,$64$)} & \makecell[l]{ResNet-$18$,\\ FC($512$,$64$),\\Linear($64$,$16$)} \\
    \hline
    $h(\cdot)$ & \makecell[l]{\textit{CSI $\&$ Vision:}\\ Linear($64$,$16$)} & / & / \\
    \hline
    $f_\text{Predict}(\cdot)$ & \makecell[l]{Linear($32$,$32$),\\ ReLU,\\ Linear($32$,$8$),\\ ReLU,\\ Linear($8$,$3$)} & \makecell[l]{AdaptAvgPool$2$d($1$,$1$),\\ Linear($64$,$32$),\\Linear($32$,$3$)} & \makecell[l]{Linear($16$,$32$),\\ ReLU,\\ Linear($32$,$8$),\\ ReLU,\\ Linear($8$,$3$)} \\
    \hline
    \end{tabular}
    \label{tab:modelconfig}
    \end{table}

    \subsection{Simulation Results}
    The performance of different methods is evaluated using $3$D Euclidean distance error. Cumulative distribution functions (CDFs) serve as the primary metric for statistical comparison of localization performance. We also validate the error correlation of single modal methods to demonstrate the complementarity of multimodal approach.
    
    \begin{figure}[t]
        \centering
        \captionsetup{font=footnotesize}
        \includegraphics[width=0.9\linewidth]{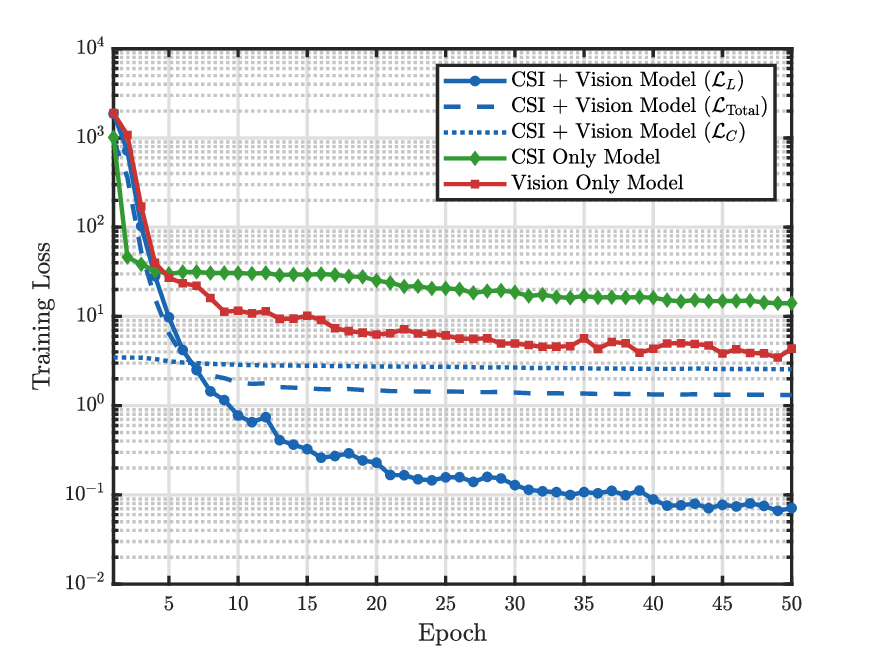}
        \caption{The curves of the training loss for different models.}
        \label{fig:loss}
    \end{figure}
    \begin{figure}[t]
        \centering
        \captionsetup{font=footnotesize}
        \includegraphics[width=0.9\linewidth]{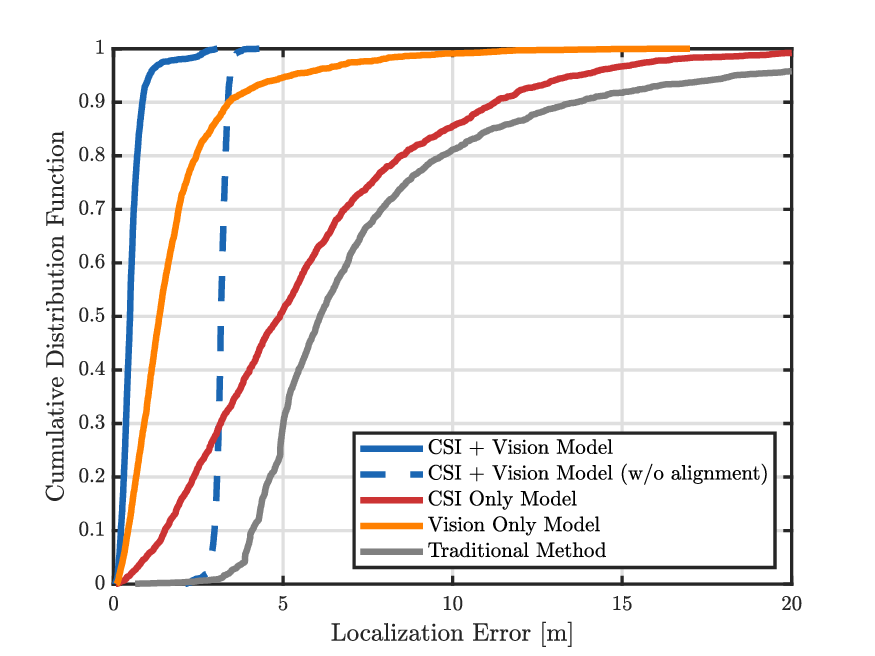}
        \caption{CDF of the localization error for different methods.}
        \label{fig:CDF}
    \end{figure}
    \begin{figure}[t]
        \centering
        \captionsetup{font=footnotesize}
        \includegraphics[width=0.9\linewidth]{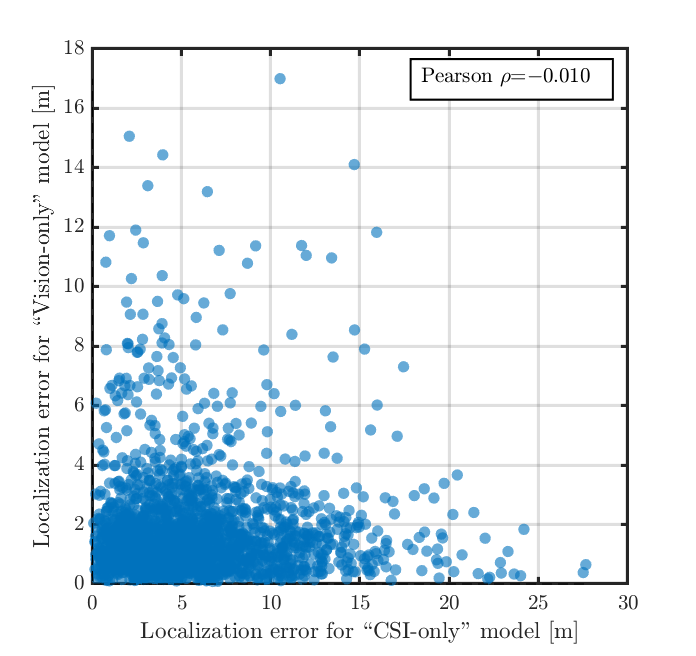}
        \caption{Localization error scatter point for single modal methods, demonstrating the correlation between the two methods.}
        \label{fig:corr}
    \end{figure}


    \subsubsection{Training Loss Analysis}
    Fig. \ref{fig:loss} illustrates the training loss curves across several baselines, multiple losses of the ``CSI+Vision" model over $50$ epochs. The localization loss $\mathcal{L}_L$ component of the ``CSI+Vision" model converges quickly below $0.1$, indicating effective position estimation. Concurrently, the contrastive loss $\mathcal{L}_C$ converges by learning to create a discriminative feature space where similar samples are clustered and dissimilar ones are separated. The adaptive weighting ensures optimal balance between these two objectives throughout training. This synergistic optimization of both direct localization accuracy and robust feature representation underpins the ``CSI+Vision" model's superior overall performance, as seen in subsequent performance analyses. In contrast, the ``CSI Only" model and ``Vision Only" model exhibit higher asymptotic loss values. 
    
    \subsubsection{Localization Error Analysis}
    Fig. \ref{fig:CDF} presents the CDF of the localization error performance of five methods. The ``CSI + Vision" model consistently demonstrates a marked advantage, achieving the highest localization accuracy. Quantitatively, it yields the lowest mean error of $0.55$ meters and a remarkably precise distribution, with $95\%$ of errors falling within $1.06$ meters. Conversely, the traditional method exhibits the least favorable performance, characterized by the highest mean error ($9.09$ meters) and a broad error spread, as its CDF curve is shifted farthest to the right. The two single modal methods provide intermediate results. The ``Vision only" model generally surpasses the ``CSI only" model in terms of mean error ($1.82$ meters versus $5.74$ meters) and at lower error percentiles. However, neither single-modality approach matches the accuracy or robustness of the multimodal method. The ``CSI + Vision" model without alignment benefits from modality fusion, resulting in a concentrated error distribution with no outliers, but its accuracy is lower, with a mean localization error of $3.17$ meters. This underscores that modality fusion ensures robustness by eliminating outliers, while alignment is critical for further enhancing performance by ensuring better feature coherence.

    

    \subsection{Complementarity of CSI and Vision Modalities}

    Fig. \ref{fig:corr} illustrates the relationship between the localization errors of the ``CSI only" and ``Vision only" models. To assess the necessity of multimodal fusion, we analyze the correlation between the localization errors of the unimodal models. Specifically, let $\boldsymbol{\Delta} \mathbf{p}_{\text{CO}}\in\mathbb{R}^{N_\text{test}}$ and $\boldsymbol{\Delta }\mathbf{p}_{\text{VO}}\in\mathbb{R}^{N_\text{test}}$ denote the localization error of the ``CSI only" and ``Vision only" models over the test set of size $N_\text{test}$, respectively. We compute the Pearson correlation coefficient $\rho$ between these two error vectors as:
    \begin{align}
        \rho = \frac{\mathrm{Cov}(\boldsymbol{\Delta} \mathbf{p}_{\text{CO}},\boldsymbol{\Delta }\mathbf{p}_{\text{VO}})}{\sigma_\text{CO}\sigma_\text{VO}},
    \end{align}
    where $\sigma_{\text{CO}}$, $\sigma_{\text{VO}}$ are the standard deviations of $\boldsymbol{\Delta }\mathbf{p}_{\text{CO}}$ and $\boldsymbol{\Delta }\mathbf{p}_{\text{VO}}$. The Pearson correlation coefficient $\rho=-0.01$ shows a negligible linear relationship between the localization errors of the two models, indicating uncorrelated, complementary error patterns. This supports fusing the modalities to potentially enhance overall localization accuracy.

    \subsection{Complexity Analysis}
    Table \ref{tab:complexity} compares model complexity in terms of network parameters and computational cost measured by Giga floating point operations per second (GFLOPS). The ``CSI+Vision" and ``Vision only" models both exhibit high complexity, with over $11$ million parameters and $1.82$ GFLOPS. This suggests that the integration of the vision component is the primary driver of the large parameter count and computational load. In stark contrast, the ``CSI only" model is remarkably lightweight, with its parameter count more than two orders of magnitude smaller and GFLOPS significantly lower. The traditional method presents the lowest computational footprint at approximately $0.001$ GFLOPS. However, the parallel processing capabilities of modern hardware (especially graphic processing units, GPUs) and software optimization have largely compensated for this disadvantage, enabling neural networks to run at acceptable speeds in practical applications. This analysis underscores the inherent trade-off between model complexity and performance, where the superior accuracy of the ``CSI+Vision" model comes at the cost of  increased computational resources compared to ``CSI-only" or traditional approaches.

\section{Conclusion}
    This paper has presented a multimodal contrastive learning regression based robust localization solution for V$2$I communication systems by fusing CSI and vision data, obtained from a single RSU and a single camera. The proposed model has significantly improved localization accuracy, achieving the lowest error. The integration of visual and CSI data has enabled better adaptation to GPS-denied conditions, ensuring high precision even in NLOS scenarios. While the multimodal approach does incur higher computational costs, its performance justifies the trade-off, making it a promising solution for future V$2$I systems in smart cities and autonomous driving applications. For future work, we plan to integrate additional modalities, such as radar data, to further enhance the robustness and accuracy of the localization system. We also aim to investigate more complex and challenging scenarios, including highly congested road conditions, environments with a larger number of users, and diverse urban landscapes. Furthermore, extending our framework to address location tracking, which involves estimating and predicting the continuous trajectory of the user equipment, will be a valuable direction for future research.

    \begin{table}[t]
        \centering
        \caption{Comparison of the network parameters and the inference cost per test sample.}
        \small
        \setlength{\tabcolsep}{6pt}
        \renewcommand{\arraystretch}{1.2} 
        \begin{tabularx}{\linewidth}{l X c c}
            \toprule
            \textbf{Model} 
            & \begin{tabular}[c]{@{}c@{}} \# \textbf{Parameters}\end{tabular} 
            & \begin{tabular}[c]{@{}c@{}}\textbf{GFLOPS}\end{tabular} \\
            \midrule
                 CSI$+$Vision             & $11,264,611$ & $1.82$  \\
                 CSI only                 & $152,963$ &  $0.0081$\\
                 Vision only              & $11,211,219$ & $1.82$  \\
                 Traditional Method       & / & $\sim 0.001$\\
            \bottomrule
        \end{tabularx}
        \label{tab:complexity}
    \end{table}
\balance   
\bibliographystyle{IEEEtran}
\bibliography{IEEEabrv, ref}

\end{document}